\documentclass{article}

\usepackage{amsfonts}
\usepackage{graphicx}
\usepackage{graphics}

\newcommand{\R}{\mathbb{R}}
\newcommand{\Z}{\mathbb{Z}}

\newcommand{\Q}{\mathbb{Q}}

\newcommand{\argmax}{\arg\!\max}

\begin{document}

\title{Sparse p-Adic Data Coding for Computationally Efficient and Effective Big 
Data Analytics}
\author{Fionn Murtagh, Department of Computing and Mathematics, \\
University of Derby, Derby DE22 1GB, UK; and \\
 Department of Computing, Goldsmiths, University of London, \\
London SE14 6NW, UK.  Email: fmurtagh@acm.org}

\maketitle

\begin{abstract}
We develop the theory and practical implementation of p-adic sparse coding of data.
Rather than the standard, sparsifying criterion that uses the $L_0$ pseudo-norm,
we use the p-adic norm.   We require that the hierarchy or tree be node-ranked, as 
is standard practice in agglomerative and other hierarchical clustering, but not 
necessarily with decision trees.  In order to structure the data, all   
computational processing 
operations are direct reading of the data, or are bounded by a constant number of 
direct readings of the data, implying linear computational time.  Through p-adic 
sparse data coding, efficient storage results, and for bounded p-adic norm stored
data, search and retrieval are constant time operations.  Examples show the effectiveness
of this new approach to content-driven encoding and displaying of data.    
\end{abstract}

\noindent 
{\bf Keywords:} big data, p-adic numbers, ultrametric topology, hierarchical clustering,
binary rooted tree, computational and storage complexity 

\section{Introduction}
\label{intro}

We start with a description of the context for this work. In \cite{ecda2014},
we provide background on (1) taking high dimensional data into a 
consensus random projection, and then (2) endowing the projected values with the Baire
metric, which is simultaneously an ultrametric.  The resulting regular 10-way tree
is a divisive hierarchical clustering.   Any hierarchical clustering can be 
considered as an ultrametric topology on the objects that are clustered.  
An ultrametric topology also expresses an r-adic number representation.  We require
r integer, $r \geq 2$.    
A 10-way tree, derived from decimal numbers, can be 
considered as an r-adic number visualization, where $r = 10$.  
In \cite{conmurt} we discussed a number of applications, and accompanying 
experimental evaluation.  How our random projection, based on uniformly distributed
random axes, differs from other work on random projection that requires Gaussian axes, is 
discussed in \cite{slds2015}.  
In \cite{ecda2015}, the consensus random projection was related to the principal 
eigenvector, addressed also was how the consensus projection is a seriation with 
clustering properties, 
and this process of seriation and clustering is closely related to the methods of 
spectral clustering, and power iteration clustering.  This is extended in \cite{ifcs2015}
with particular reference to Correspondence Analysis.

In \cite{ecda2014, slds2015}, the context for the use of the following data is described:
34,352 research funding proposals indexed in the open source Apache Solr package for 
(server-side) indexing, and (client-side) querying, search and retrieval.  This package 
supports search and retrieval in large document collections, consisting of various fields 
(such as title, authors, abstract, and any other field that is defined for the 
document collection).  The Solr packages determines and uses a similarity 
between documents that it calls MLT (``more like this'').  This similarity has 
fixed field weights, and otherwise weights the document/term indexed data (see
further discussion in \cite{murtaghmlt}).  We used scores generated by 
Solr for the top 100 matching proposals for each of a selection of 10,317 of the proposals 
set. 

\begin{figure}
\centering
\includegraphics[width=12cm]{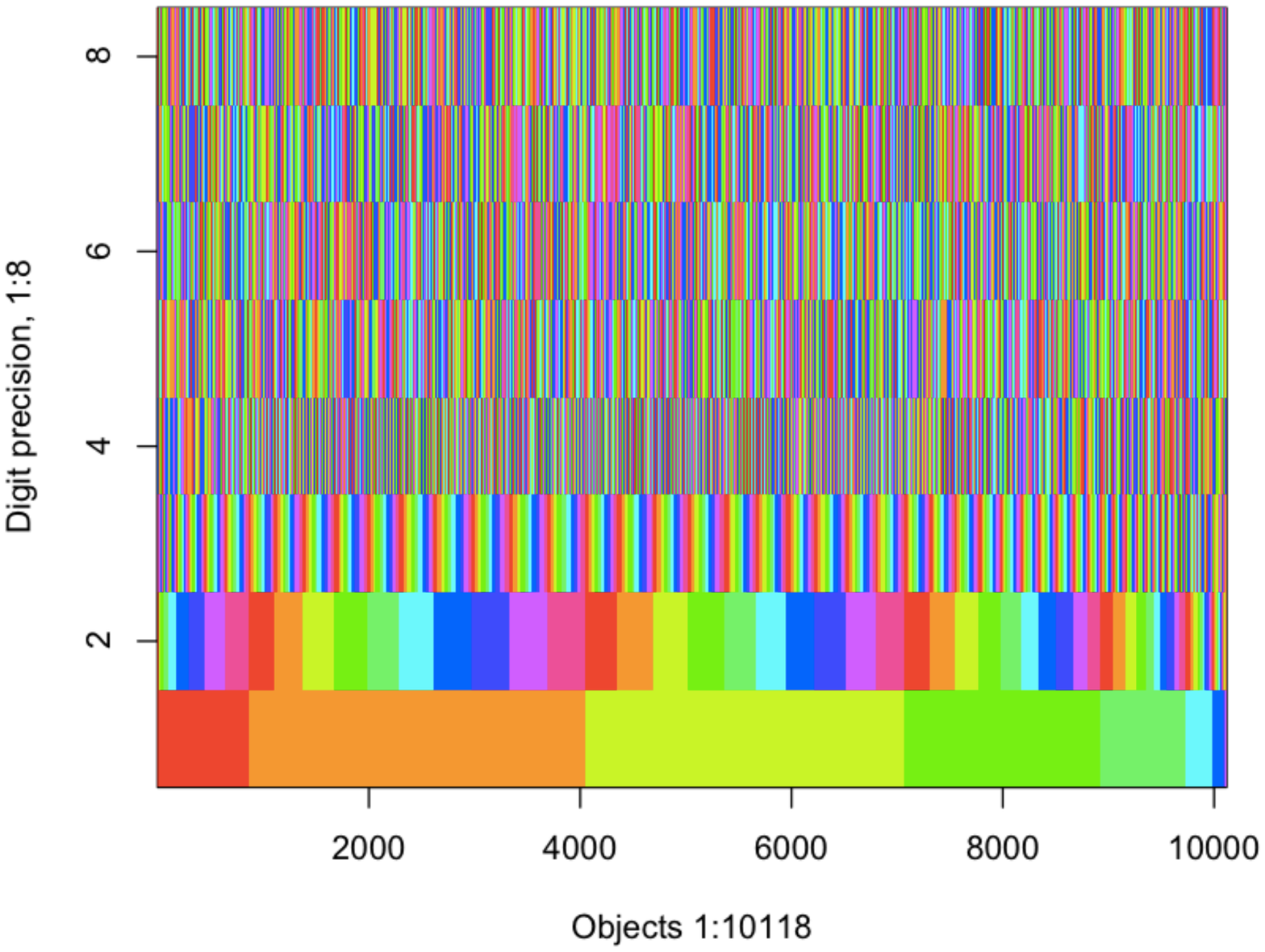}   
\caption{Means of 99 random projections.  Abscissa: the 10118 (non-empty)
documents are sorted (by random projection value).   Ordinate: each 
of 8 digits comprising random projection values. A rainbow colour coding is used
for display.}
\label{fig10}
\end{figure}

Using a regular 10-way tree, Figure \ref{fig10} shows the hierarchy produced,
with nodes colour-coded (a rainbow 10-colour lookup table was used), and with the root 
(a single colour, were it shown), comprising all clusters, to the bottom.  The 
terminals of the 8-level tree are at the top.  

This divisive hierarchical clustering algorithm uses the Baire, or longest 
common prefix, distance, which is also an ultrametric, on the consensus random 
projection values.   

While this Figure \ref{fig10} is an unorthodox display of a dendrogram, or 
hierarchical clustering, its role as a display is used here for exposition
and discussion, and it is better depicted in this way, compared to a regular
10-way tree.   

The first Baire layer of clusters, displayed as the bottom level in Figure \ref{fig10}, 
was found to have 10 clusters (6 of which are very evident, visually.) The next Baire 
layer has 87 clusters (the maximum possible for this 10-way tree is 100), and the third 
Baire layer has 671 clusters (maximum possible: 1000).  

The Baire hierarchy has a limitation in regard to storage.  We need to allow
for a regular r-way tree, here a 10-way tree, for storage.  For finding a path 
in a hierarchical descent from the root of the tree, traversal length is logarithmic
in the number of objects clustered, i.e.\ associated with the terminal nodes.  However
storage of all nodes in the tree, when the precision and hence the number of digits per
object is not fixed, is exponential.   

We propose the following solution in order to linearize storage.  First, we 
invoke the natural sparse coding of a p-adic ($p$ prime) or more general 
r-adic ($r$ integer) representation.  We require checking of the general class of 
data that is at issue, in order to limit the precision.   Secondly, we reduce the 
base $r$ of the number representation used, in order to further economize on Baire 
tree information. 
We do this by means of a best approximation of the data in the new base, 
relative to the data in the former base. 

Motivation for this work includes the varied perspectives that are at issue in the 
following: \cite{ref444, dragovich, hall, khrenn2, mirkin97}.

This paper is structured as follows.  

In \cite{refmur90a}, some of the ways are reviewed that we can p-adically encode 
a hierarchical clustering, a classification tree, that is conventionally a rooted, 
binary, labelled and ranked tree.

In section \ref{introsparse} we introduce sparse coding. General motivation for r-adic 
and p-adic (r integer, p prime) data coding is provided in section \ref{motsparse}. 
Then specification is provided in section \ref{radic}.  
Our first method for implementing 
p-adic data encoding is set out in section \ref{algsect}.  In section \ref{repapprox}
a further algorithm is proposed of the p-adic or m-adic ($p, m < r$ where r-adic
is the starting context) representation.

In this work we are not just dealing with a tree structuring, viz.\ a hierarchical 
clustering, followed by encoding, but rather a tree structuring that represents the 
data.  In this particular way, data content is taken into account.  In section 
\ref{repapprox}, with the aim of storage economy, we do take the hierarchic 
representation alone, and simplify that, through p-adic or m-adic mapping (p prime, 
m an integer, with p or m $<$ r, r the integer that is used 
in our initial r-adic representation).  

\section{Introduction to Sparse Coding}
\label{introsparse}

Sparse coding is used to support data compression.  Additionally compressive 
sampling is the use of appropriate sparse coding in order to support data 
reconstruction.  Widely used examples of sparse coding are run length encoding, 
use of orthonormal mapping through eigen-reduction, and the use of 
sparsifying signal transforms including the Fourier, wavelet, discrete cosine,
and other multiple resolution transforms.  In this work, we demonstrate how p-adic
number representation provides a most advantageous way to sparsely encode big data sets.

A short introduction to sparse coding is \cite{ng-sparse}.  The $i$th input vector 
$x_{iJ} \in \R^J$ is written as a linear combination of a possibly over-complete 
(i.e., $K > J$) basis, $\phi$, with each $a_{ik}$ a scalar coefficient: 
$x_{iJ} = \sum_{k=1}^K a_{ik} \phi_{kJ} = a_{iK} \phi_{KJ} $.  Due to 
over-determination of the problem of estimating the coefficients, a 
sparsity cost function is also taken into 
consideration.  A standard objective for such sparse coding is
as follows (where we are rewriting in matrix notation, and replacing norm with distance):

$$ \min_{a_{iK}, \phi_{KJ}} 
\sum_{i=1}^I  \| x_{iJ} - \sum_{k=1}^K a_{ik} \phi_{kJ} \|_2^2 + \lambda \| a_{iK} \|_0 $$
$$ = \min_{a_{iK}, \phi_{KJ}} 
\sum_{i=1}^I  \| x_{iJ} - a_{iK} \cdot  \phi_{KJ} \|_2^2 + \lambda \| a_{iK} \|_0   $$
\begin{equation}
= \min_{a_{iK}, \phi_{KJ}}
\sum_{i=1}^I d_2 ( x_{iJ} - a_{iK} \cdot  \phi_{KJ} )^2 + \lambda \| a_{iK} \|_0
\label{sparsecoding}
\end{equation}

\noindent
i.e., minimize in a least squares sense, using Euclidean, $L_2$, 
distance, the separation of 
$x_{iJ}$ from its approximation by the linear combination of $\phi_k$ terms, 
$1 \leq k \leq K$, 
each term weighted by $a_{ik}$; and that minimization subject to the $L_0$ pseudo-norm, 
implying that the number 
of non-zero terms is to be minimized.  For this optimization, $\lambda$ is a Lagrange
multiplier.  

Due to its differentiability and convex optimization to provide a solution 
for it, $\| a_{iK} \|_0$ is generally replaced, as an approximation, 
by $\| a_{iK} \|_1$.   Since the $a_{ik}$ terms may decrease 
in value, but then the $\phi_k$ terms could be large, an additional constraint 
can be applied: $ \| \phi_k \| \leq C$, for all $k$, and for constant $C $.

We restate the foregoing by saying that we want to represent all our $x_{iJ}$ data 
vectors 
by a basis expansion, the $\phi_k$ being the basis, and having $K$ coefficients, also 
termed loadings or projections.  The coefficients are the scalar product, 
$a_i^t \phi$, of these two $K$-valued vectors.  Our aim is to have a small number of 
non-zero coefficients in the vector of coefficients, $a_{iK}$.  

As discussed briefly in \cite{ng-sparse}, encoding a new vector $x_{iJ}$ typically 
requires re-optimizing the vector of coefficients, $a_{iK}$.

In general applications, the significant coefficients are of most interest.   The 
signal is made more sparse by using some compactifying transform (such as a wavelet 
or other multiresolution transform).  Then, in transform space, small-valued 
transform coefficients are thresholded.  This also can be applied for noise filtering,
using hard or soft thresholding of wavelet 
coefficients.  (That is, setting wavelet coefficients to zero if they are less than, 
for example, a statistically determined significance threshold, and then subsequently 
applying the inverse wavelet transform in order to reconstruct the data.)  In 
compressive sensing \cite{hayashi}, 
(1) the data is sampled through a sensing protocol; and then 
(2) the basis to be determined, $\phi$ above, is the product of the sensing protocol 
matrix and 
appropriate dictionary atoms; (3) since in a multiresolution transform dictionary such as
wavelets, the coefficients have a power law decay, the estimation of these coefficients, 
$a_i$ above, retains a set number of the largest entries.  One arrives at a 
$c$-sparse encoding, such that $\| a_{iK} \|_0 \leq c $ (i.e.\ $ \leq c$ non-zero
coefficients retained).

Our new perspectives include the following.  

\begin{enumerate}

\item Instead of a signal compactifying 
transform (such as a wavelet or other multiresolution transform), we use a tree or 
ultrametric topology in which our data is embedded.  The partial order needs to 
take fully into account the importance or significance properties of the data used.
We achieve this objective through p-adically encoding our data.  (In \cite{baraniuk},
wavelet transform compactifying makes use of sets of branchings connecting wavelet
scales. This then takes account of e.g.\ image or signal edges showing up as a 
succession of high wavelet coefficients along the branch of a wavelet tree structuring 
of the image; and the alternative, low wavelet coefficient values, in the case
of smooth regions.)  

Our embedding in an ultrametric topology, expressed
p-adically, is, by design, compactifying.  That is, we take account of the given 
proximity (or metric) properties of our data.  

\item For $c$-sparse encoding, hence minimizing 
$\| a_{iK} \|_0$, we will instead implement this 
constraint using the p-adic norm.
 
\item To address the issue raised in \cite{ng-sparse}, whereby new data implies 
re-optimization, we will require the following: the ultrametric tree topology 
corresponding to our p-adic representation is a {\em regular} tree topology; also, 
this tree topology is a rooted, {\em ranked}, tree, viz.\ for each node in the tree,
there is a real-valued level value.   

\end{enumerate}


\section{Motivation for p-Adic Sparse Coding}
\label{motsparse}

p-Adic numbers are endowed with a very different order structure compared to real 
numbers.   Following a short introduction to p-adic numbers, our objective in this 
work is to exploit the natural (topological rather than Hilbert space geometric) 
ordering of p-adic numbers in sparse coding.  

p-Adic numbers were introduced by Kurt Hensel in 1898.  
The p-adic numbers are base p numbers, where p is a prime number.  The reals are 
expressed in terms of a p-adic number systems where p is infinity.
The ultrametric
topology was introduced by Marc Krasner in 1944 \cite{krasner}, 
the ultrametric inequality having been formulated by Hausdorff in 1934. 
 Essential motivation for the study of this area is
provided by Schikhof \cite{schikhof} as follows.  Real and complex fields gave rise
to the idea of studying any field $K$ with a complete valuation $| . |$
comparable to the absolute value function.  Such fields satisfy the
``strong triangle inequality'' $| x + y | \leq \mbox{max} ( | x |,
| y | )$.  Given a valued field, defining a totally ordered Abelian group,
an ultrametric space is induced through $| x - y | = d(x, y)$.
Various terms are used interchangeably for analysis in and
over such fields such as p-adic, ultrametric, non-Archimedean, and isosceles.
The natural geometric ordering of metric valuations is on the real line,
whereas in the ultrametric case the natural ordering is a hierarchical
tree.  p-Adic numbers,
which provide an analytic version of ultrametric topologies,
have a crucially important property resulting from Ostrowski's theorem.
Each non-trivial valuation on the field of the rational numbers is equivalent
either to the absolute value function or to some p-adic valuation
(\cite{schikhof}, p.\ 22).  Essentially this theorem states that the rationals can be
expressed in terms of (continuous) reals, or (discrete) p-adic numbers, and
no other alternative system.

A generalization of integer coding, as well as Huffman (prefix) and Golomb-Rice 
(two-symbol alphabet) coding, 
is studied in \cite{rodionov}.  While being acknowledged as being very close in 
operation to these widely used entropy coding algorithms, in \cite{rodionov2} the
authors point to how all are special cases of one p-adic coding algorithm.  Many 
examples are provided in \cite{rodionov2}, using implementation in the Ruby programming 
language.  

In \cite{bradley}, the so-called split-LBG clustering algorithm is considered in a 
p-adic framework.  This clustering algorithm, \cite{LBG}, is used for data quantization.  
Motivation for this work is dynamic contexts, where cluster centres and data elements
considered, can change over time.   Another top-down, or divisive, hierarchical
clustering is proposed in \cite{mirkin03}.  Our work, that uses the Baire hierarchy,
seeks digit, and hence integer, representatives of clusters or nodes in the tree. In 
adopting such an approach, we very easily arrive at p-adic coding.   

\section{Approximation of r-Adic (r Integer) Representation}
\label{radic}

We proceed as follows.
Express our data as r-adic numbers.  Take a usual decimal representation of real valued
numbers, to finite precision, base $r = 10$ and precision is the $c$th integer
value.  

A direct expansion of $x = \sum_{k=0}^c a_k p^k$ offers no guarantee of a smaller,
resulting number of digits, compared to the original base, r.  
Expansion can be considered for $p < r$ or $p > r$, and
for representation, digit-wise, of the given $x$, base $r$, and the full number.    
Given our use of coding to expedite search and retrieval, we therefore determine close
approximation to our data.  This is analogous to a transform coding, referred 
to in section \ref{introsparse}, used subsequently for sparse coding. 


Rewriting a 
$c$-digit, base $r= 10$, number in some other base, $x = \sum_{k=0}^c a_k p^k$, 
is not of benefit, because of computation required (more so than the stages of our 
Baire-based algorithm which, by design, has linear time requirements), and also due to 
the undetermined digits of precision, and hence overall storage, required.  

Motivated by the problem specification of equations (\ref{sparsecoding}), we ask what 
benefits could there be if, instead of the $L_2$, together with $L_0$ and $L_1$ metrics, 
we use the Baire metric, which is simultaneously an ultrametric.  See 
\cite{conmurt,murtcon}, the former for applications, and the latter for roles in 
computational science.   For vectors $x_{iJ}, x_{i'J}$, consider a basis 
$r$, which can be a prime, notationally $p$, or non-prime, for example, decimal, $r = 10$.
Definition of the Baire metric: $d_r (x_{iJ}, x_{i'J}) = r^{-\beta} \mbox{ such that } 
\beta = \argmax_j \{ x_{ij} = x_{i'j} | 1 \leq j \leq J \}$.  

We seek to minimize 
$$
\sum_{i=1}^I d_2^2 ( x_{iJ} - a_{iK} \cdot  \phi_{KJ} ) $$
\begin{equation}
= \sum_{i=1}^I \sum_{j=1}^J \left( x_{ij} - \sum_{k=0}^{c-1} a_{ik} \phi_j^k \right)^2 
\mbox{ with } \| x_i \|_r \leq r^{-c} \mbox{ and } a_{ik} < r
\label{sparsecoding2}
\end{equation}

In seeking this best approximation, given ordering of r-adic expansion 
through the 
$\phi^k$ term, this natural ordering allows, for the r-adic digits, the best 
1-, 2-, ... $c$-approximation to be derived from the r-adic representation.  

Informally expressed, we have the following.  At each level of the regular
tree, we approximate the values that we have at that level by a best match to 
all of these values.  Also the number of levels in the regular tree is bounded
by design. 

\section{p-Adic Encoding Algorithm Description and Implementation}
\label{algsect}

Given data in a space of any dimensionality, (1) we apply a consensus of random projections,
and (2) we induce a Baire hierarchical clustering on these projections.  Given that the 
projections are of digit precision $J$, in a real number and hence decimal by default 
system (i.e.\ $r = 10$), the projected values are endowed with the Baire distance (and
ultrametric).  This Baire distance is the longest common prefix, of the digits, 
ordered by precision.   

We can display the Baire hierarchy, for $I$ data points, and for digit precision $J$, 
as an $I \times J$ array.  Each value in this array is a base $r = 10$ value.  We have: 
$A: I \times J \rightarrow \{ 0 , 1, 2, \dots , r-1 \}$ where $r = 10$.   Consider any 
digit of precision, $j_\alpha$.  Fix the value of $j_\alpha \in \{ 0, 1, \dots , r-1 \}$.
Consider all child nodes of this node.  Unless any node is empty, there are $r$ child 
nodes of this selected node.  Cf.\ Figure \ref{fig10} and later figures also.

In \cite{conmurt} we noted how digits at differing precision levels display different
distributions.  In that work we noted how different digit distributions can be used 
as a novel discriminating feature, and we exemplified this on astronomical spectrometric
and photometric redshift values.  This was in the context of nearest neighbour regression.
A consequence for this work in this paper is that we will approximate our given $A$ 
values independently by precision digit. $A$ is our array display of the hierarchy.

More generally, patterns and, indeed, anomaly in data digits 
can be of major benefit in, e.g., forensic data analytics.  Cf.\ Benford's law
\cite{benford,hill,berger}. 

We map the base $r = 10$ digit values onto a base $p$ system.  We examined  
$p = 7, 5, 3, 2$, and we report below on some of these.  
It is not a requirement that $p$ be prime.  What does follow from 
our algorithm is that the closer $p$ or an alternative target number base is to $r = 10$, 
then the better the approximation, that we will form to $A$, will be.

In order not to overload our notation, we will consider any one given digit of precision,
$j = 1, 2, \dots , J$.  See Figure \ref{fig10}, where we are taking into consideration
at any given time, one row, or one level of digit precision.
We do this in view of the observed different distributional characteristics of digits 
that we reported in \cite{conmurt}, and in view of Benford's law.  Were it the case of 
compression as our sole objective, then we would take all digits into consideration, hence
all $j$.  

From equation (\ref{sparsecoding2}), we seek to determine the minimum of:  

\begin{equation}
( x_i - \sum_{k=0}^{K-1} a_{ik} \phi^k )^2
\label{sparsecoding3}
\end{equation}

for all objects, $i$, hence, we seek 
to determine the minimum of:

\begin{equation}
\sum_{i=1}^I ( x_i - \sum_{k=0}^{K-1} a_{ik} \phi^k )^2
\label{sparsecoding4}
\end{equation}    

Write $x_i$ by digit of precision, $x_{ij}$.  Further we specify that the set of objects,
$i$, must be in a partition, to be determined, of the object set.  Call this partition 
of the objects, $o_0, o_1, \dots o_{K-1}$.   
Alernatively expressed, for some $k$, each index $ i \in o_k$.   
So, having $K-1$ values at this, or any, level, then the partition class of value
$0 \leq k \leq K-1$ is $o_k$. 
Following this additive decomposition (by partition) of $x_i$, we replace equation
(\ref{sparsecoding4}) by equation (\ref{eqn100}):

\begin{equation}
\sum_{k=0}^{K-1} \sum_{i=1}^I ( x_{io} - a_{ik} \phi^k )^2 \mbox{  with  } o = o_k 
\label{eqn100}
\end{equation}

(We write $o = o_k$ here solely for notational clarity.)
The first term in this expression is a real value, and the second term is a $K$-adic 
value.  If p-adic, then this second term is in the field of p-adic numbers, $\Q_p$.
Therefore as such, we must re-map the rightmost term into the reals.  In order 
to do this, consider what is done in quantization for compression tasks: codewords are 
used, assembled in a codebook, and accessed as a function of the given data values.  This 
we do now as follows: for all $i = 1, 2, \dots , I$, 
$a_k \phi^k \rightarrow c_k$, with $c_k \in \R$.  Thus, equation (\ref{eqn100}), which we seek 
to optimize through choice of codewords, $c_k$, and partition of objects, $o_k$, for $k =
0, 1, \dots , K-1$, becomes:
\begin{equation}
\sum_{k=0}^{K-1} \sum_{i=1}^I ( x_{io} - c_{ik} )^2 \mbox{  with  } o = o_k 
\label{eqn101}
\end{equation}

This is the classical K-means clustering problem where we seek clusters of the object set,
the union of these clusters is the object set, and the codewords are the cluster centres 
or centroids.  Iterative optimization, from random initialization of cluster centres,
provides the solution.  See e.g.\ \cite{bock}.  

For completeness of exposition, we can return to the levels $j$ corresponding to digit 
precision, in equations (\ref{eqn100}), (\ref{eqn101}): 

\begin{equation}
 \sum_{k=0}^{K-1} \sum_{i=1}^I ( x_{ijo} - c_{ijk} )^2 \mbox{  with  } o = o_k, 1 \leq j 
\leq J 
\label{eqn102}
\end{equation}
Thus:
\begin{equation}
\sum_{k=0}^{K-1} \sum_{i=1}^I ( x_{ijo} - a_{ijk} \phi^k )^2 \mbox{  with  } o = o_k,
1 \leq j \leq J
\label{eqn103}
\end{equation}
As we have noted above, we retain the separateness of digit precision, in view of our 
possible interest in their distributional properties.  (Also as noted above, if 
compressibility were the sole objective, then the optimization of equations 
(\ref{eqn101}), (\ref{eqn102}) or (\ref{eqn103}) 
would be summed over $j$.)

An optimal solution to equation (\ref{eqn101}) implies an optimal representation of 
our set of reals expressed as a K-valued, K-adic expansion, 
equation (\ref{eqn100}).   In practice, our iterative 
optimization of K-means in regard to the optimand, equation (\ref{eqn101}), is suboptimal.
This is simply due to K-means partitioning 
being NP-complete.  In usual operation, firstly a large 
number of iterations are permitted, and secondly, a number of different initialization 
configurations are used to provide the best overall result, or a consensus result.  (In 
our implementation, R function {\tt kmeans} was used with maximum 500 iterations and 
50 random starts.)

\begin{figure}
\centering
\includegraphics[width=12cm]{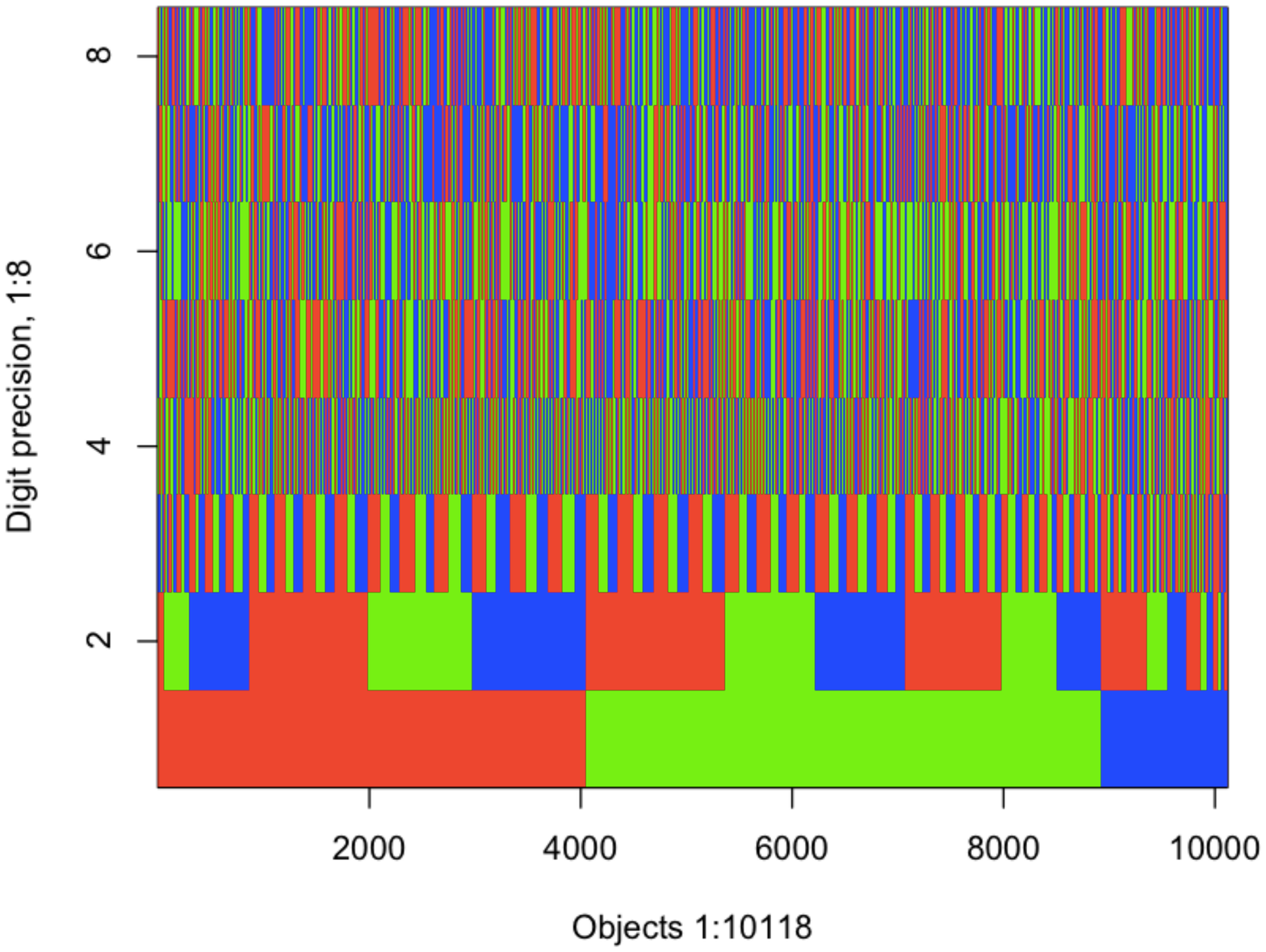}
\caption{Abscissa: the 10118 (non-empty) documents are sorted (by random projection 
value).   Ordinate: each of 8 digits comprising random projection values.  3-adic 
representation, displayed using three colours.}
\label{fig100}
\end{figure}

Figure \ref{fig100} shows the 3-adic representation that results for Figure \ref{fig10}. 
For 
computational convenience, these are digits 1, 2, 3 representing 0, 1, 2 in the 3-adic
representation.  Mean squared error (MSE) is determined from the 3-adic expansion, 
together with the codewords from the codebook.  (These are the cluster centres 
determined by the K-means algorithm.)
The MSE at digit precisions 1, 2, \dots , 8 were as follows:
0.26, 0.90, 0.88; 0.88; 0.89; 0.89; 0.90; 0.90.
The p-adic Baire tree in this figure is a regular 3-way tree or hierarchy.  

\begin{figure}
\centering
\includegraphics[width=12cm]{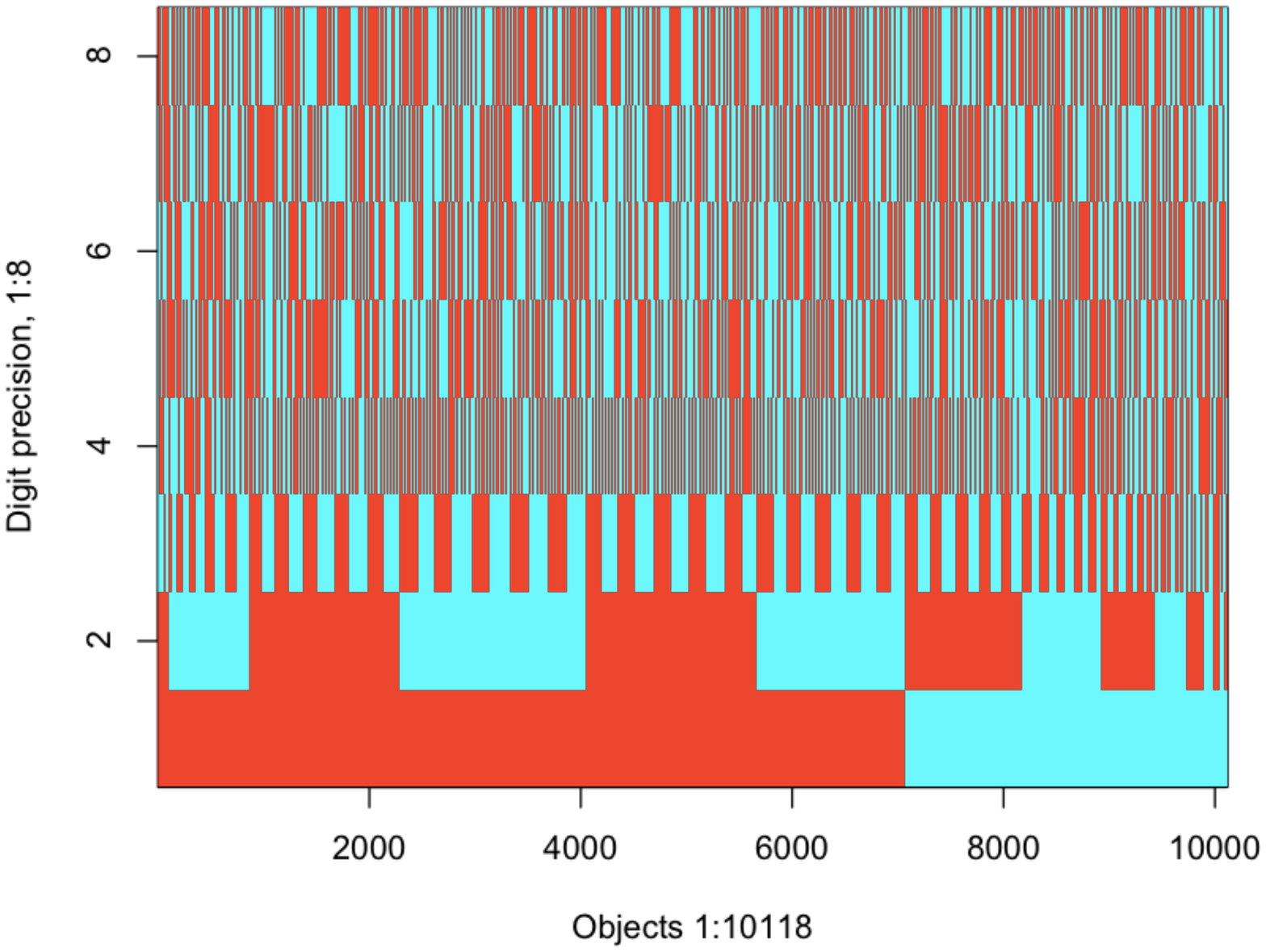}
\caption{Abscissa: the 10118 (non-empty) documents are sorted (by random projection 
value).   Ordinate: each of 8 digits comprising random projection values.  2-adic, 
or binary, representation, displayed using two colours.}
\label{fig1000}
\end{figure}

Figure \ref{fig1000} shows the 2-adic or binary representation of Figure \ref{fig10}. For 
computational convenience, these are digits 1, 2 representing 0, 1 in the 2-adic
representation.   The MSEs at digit precisions 1, 2, \dots , 8 were as follows:
0.56, 1.96, 1.98, 2.01, 1.97, 2.03, 1.97, 1.99.
The p-adic Baire tree in this figure is a regular 2-way, or binary, tree or hierarchy.  

Computationally all stages of the processing are a linear function of cardinality of 
the object set, and of the levels or digits of precision, of numbers of iterations 
in the K-means work.  Thus overall our preprocessing, i.e.\ data structuring, 
is of linear computational time. 
Through application of filtering in the sparsely encoded data, i.e.\ limiting the 
p- or r-adic expansion to a fixed number of terms (alternatively expressed: limiting
to a fixed number of digits of precision in the chosen number base), we also have 
storage that is linear in the number of objects.  

A binary tree representation (Figure \ref{fig1000}) is particularly 
appropriate for decision making
and related binary routing, and such a clustering display, determined in a top-down
or divisive way can be contrasted with an agglomerative hierarchical clustering 
algorithm.  

To check on this, the following assessment was carried out.  The mean random 
projection set, 10118 values, was hierarchically clustered, using Ward's minimum
variance method.  The input data (cf.\ Figure \ref{fig10}) were to full available 
precision, viz.\ 9 digits of precision.  
The real values representing the binary expression from Figure
\ref{fig1000} were also hierarchically clustered using Ward's method.  For the 
projections, the correlation between input distances and cophenetic 
(or ultrametric, or tree) distances determined from the hierarchical clustering
was 0.6137619.  For the binary, 2-adic hierarchy, the correlation between input 
distances (i.e.\ from the reals directly expressing the 2-adic, 8-level hierarchy),
and cophenetic distances from the Ward hierarchical clustering, was 0.7257293.
The correlation between the two sets of input distances was only 0.4872135. But 
actually the correlation between the values themselves (recalling that both 
sets of values comprised vectors of 10118 values) was 0.7102683.  Finally 
the correlation between the two sets of cophenetic distances was 0.6154983.

The correlation between the two sets of values (mean random projection, as noted, 
of 9 digits of precision, and 
its 2-adic, 8-term expansion, that we determined through original digit optimized 
fit), viz.\ 0.7102683, is our most relevant finding.  Note how levels of precision 
differ (9 versus 8), that in the former case we have digits that range over 0 to 9,
whereas in the 2-adic representation case, we have digits that range over 0 and 1 
(although, whenever convenient for implementation, 
we express them as, respectively, 1 and 2).  So in the 
former case we have to consider, for each object, 10 possible values for each of 9 
digits; and in the latter case, we have, for each object, 2 possible values for each of 
8 digits.   

For proximity matching or other such operations, we therefore require:

\begin{enumerate}
\item The random projection vector.
\item The codebook for each digit level.
\end{enumerate}

We term the foregoing approach the codebook through cluster-based representation.

\section{p-Adic and m-Adic Fitting through Approximation of Hierarchy Representation}
\label{repapprox}

We now seek to simplify the representation in the following way. Label the 
digits of precision, or layers, or levels, $j = 1, 2, \dots $, which in the case of 
the study here are of 
maximum value, 8.  The set of objects is indexed by $i$, and in the study 
here, $i = 1, 2, \dots , 10118$.  We will refer to the Baire hierarchy 
representation of objects crossed by digits, as the Baire array display, $A$,
cf.\ section \ref{algsect} above, with elements $a$.  
For digit set, $J$, and object set, $I$, we have the Baire array display
defined by the mapping: $A : I \times J \rightarrow \Z_\nu$, where $\Z_\nu$ is 
the set of integers modulo $\nu$.  In the decimal, or base 10, number system,
which is our point of departure, $\nu = 10$ and we consider the decimal 
digits, $\Z_\nu = \{ 0, 1, 2, \dots , 9 \}$.  

\begin{enumerate}
\item 
For a given digit level, $j$, and given object neighours, $i$ and $i'$, we take
(without loss of generality) $i' = i + 1$. We consider the Baire 
array display values, $a_{i, j}, a_{i', j}$.  
\item 
As a first step we determine all
neighbour pairs,  $a_{i, j}, a_{i' j}$, that have the same parent digit value.  
That is, $a_{i, j-1} =  a_{i', j-1}$.  
\item The next step in our algorithm is to 
determine if our Baire array display values differ by 1, viz.\ $| a_{i, j} - 
a_{i', j} | = 1$.  
\item If, instead, $a_{i, j} = a_{i', j}$, then no intervention is 
required.
\item If $| a_{i, j} - a_{i', j} | > 1$, then we will not intervene insofar as 
there is sufficiently clear difference between these Baire array display values.
\item We determine the number of neighbour values differing by 1. 
That is, we determine: $ {\cal N} = |  ( a_{i, j}, a_{i', j} )|, i = i^*, i' = i^* + 1, 
\mbox{  such that }
| a_{i, j} - a_{i', j} | = 1$, i.e.\ the number of these $i^*, i^*+1$ pairs. 
\item We determine the minimum such values of these pairs: 
$\mbox{argmin}_{i^*, i^*+1} {\cal N}$.  For any representative values of the indices,
$i', j$, call the larger value that realizes this, $a_{i', j} = w$.  We have 
$w \in \Z_\nu$. 
\item Update the Baire array display, $a$, as follows: 
if $k \geq w$, then for $a_{i, j} = k, \ a'_{i, j} = a_{i, j} - 1$.  
That is, for those Baire array display pairs, with the same parent digit value,
 that differ by 1 in their own digit values, we set the higher digit value to be 
equal to the lower digit value; then we make this update of those digit values 
throughout the entire Baire hierarchy representation.   
\item It results that the Baire hierarchy representation, with values in 
$\Z_\nu$, has been transformed into an optimal (resulting from step 7) approximating 
Baire array display representation, with values in $\Z_{\nu-1}$.
\item This procedure, steps 1--9, can be repeated.  On each such iteration, 
we decrease the number of values in the Baire array display by 1.
\end{enumerate}

\subsection{Example of Application}

\begin{figure}
\centering
\includegraphics[width=12cm]{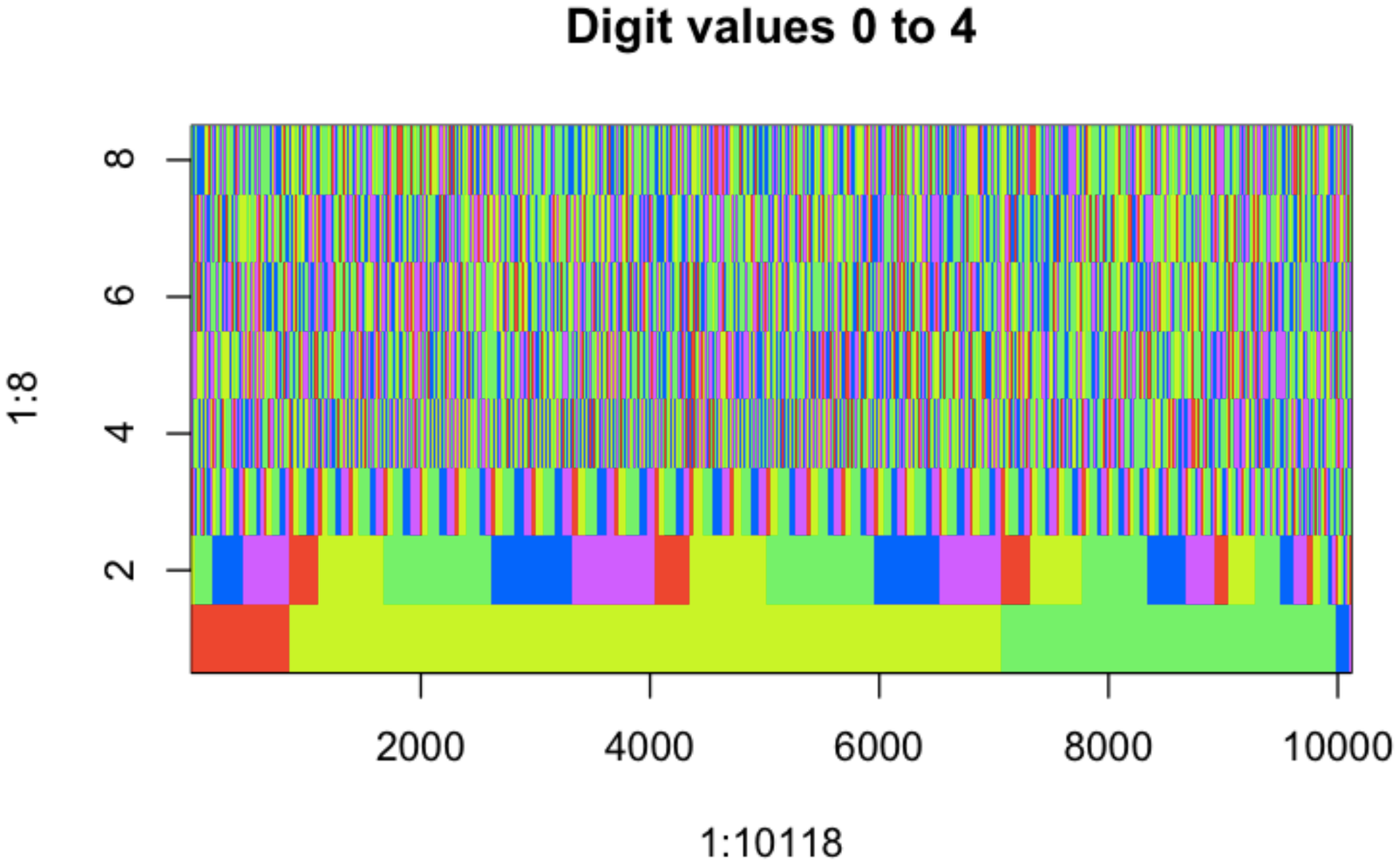}
\caption{Fifth approximation by merging close neighbour pairs of a common 
parent node.}
\label{fig20}
\end{figure}

\begin{figure}
\centering
\includegraphics[width=12cm]{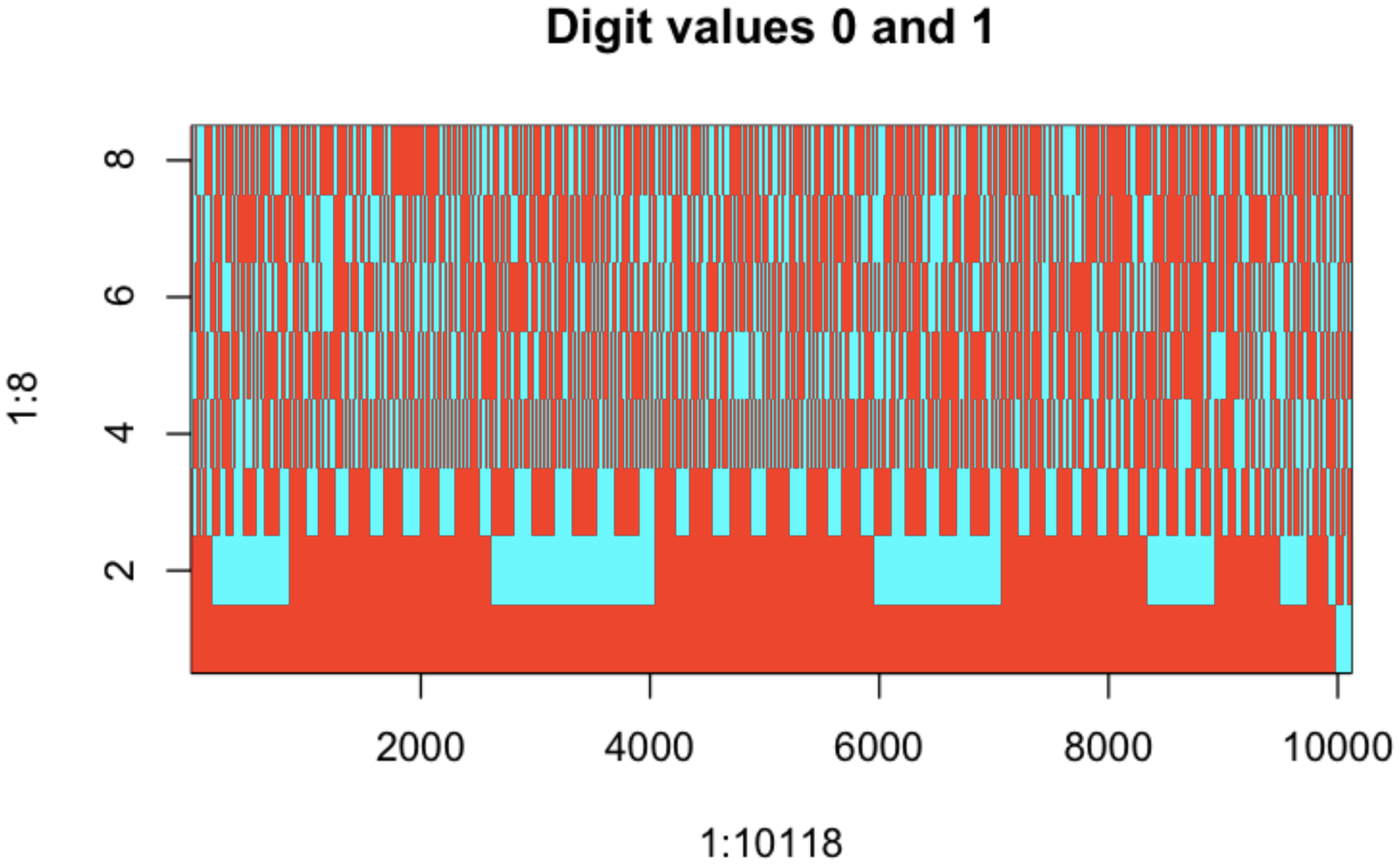}
\caption{Eighth approximation by merging close neighbour pairs of a common 
parent node.  Display of a binary tree.}
\label{fig30}
\end{figure}

We use Figure \ref{fig10}. This figure is displayed with digit values 0 (taken 
here as a value, and not the absence of a value) to 9, and 
we have it as 10-adic encoded.  The approximation algorithm described in the 
previous subsection is applied, stepwise.  The initial 10-adic representation 
is approximated by 
a 9-adic approximation.  From the 9-adic encoded hierarchy representation, 
an 8-adic encoded hierarchy representation approximates it.  Then that is 
approximated by a 7-adic encoded hierarchy representation.  Then a 6-adic 
encoding approximates that.  Then a 5-adic encoding approximates that, and
is shown in Figure \ref{fig20}.  This is followed by a 4-adic encoding, a 3-adic
encoding, and finally a 2-adic, binary, encoding.  The last one here is shown 
in Figure \ref{fig30}.

\subsection{Approximation Distance and Observed Convergence of m-Adic Approximation 
of the Baire Array Display}

\begin{figure}
\centering
\includegraphics[width=12cm]{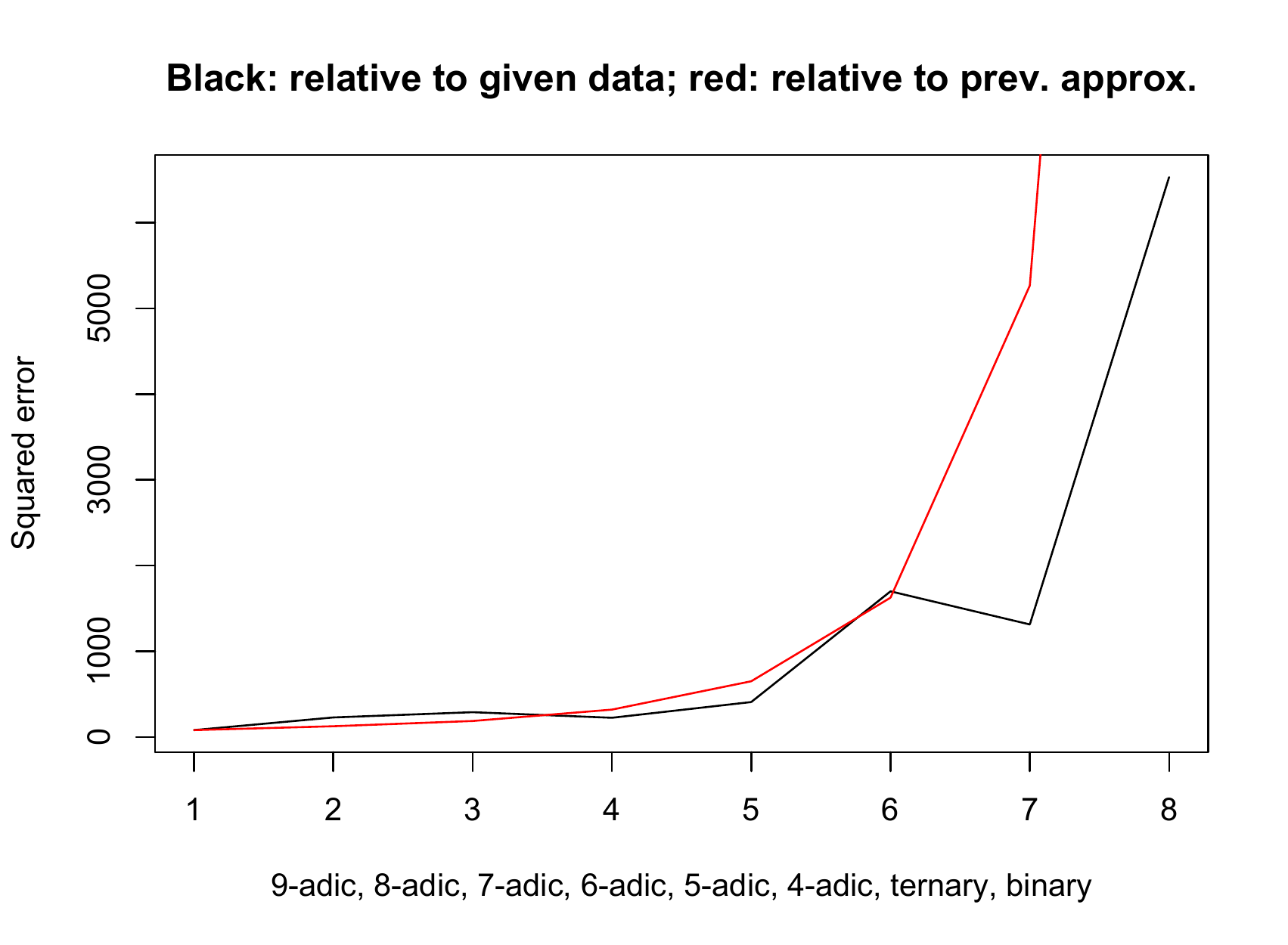}
\caption{Black curve: measure of approximation by the succession of m-adic approximations,
starting with the 10-adic Baire array display, 
proceeding with the 9-adic, 8-adic, 7-adic, 6-adic, 5-adic, 4-adic, 3-adic or 
ternary, 2-adic or binary.  Approximation is measured by squared error. 
Red curve: measure of convergence, being the difference, 
viz.\ the squared error, between the m-adic approximation and the previous approximation.
Approximations and the original Baire array display have normalized values, in [0,1].}
\label{fig40}
\end{figure}

Figure \ref{fig40} displays the squared distance, firstly, between the m-adic 
approximation and the given Baire array display.  That is, we have 10-adic or 
decimal data to start with.  See Figure \ref{fig10}.  Just two of our m-adic approximations 
are displayed here in Figures \ref{fig20} and \ref{fig30}.  In Figure \ref{fig40} 
there is the growing discrepancy between the m-adic approximation and the given decimal 
data.  

Note that the approximation to the given Baire array display, as is the latter,
are normalized, so that values are real-valued and bounded by 0 and 1.

Secondly, Figure \ref{fig40} shows the convergence properties.  For m $= 9, 8, 
\dots, 3, 2$, we consider the squared error between the m-adic approximation and
the m+1-adic approximation, or, initially, the given decimal data. 

We note from Figure \ref{fig40} that up to m = 5, which as a prime, we can write
conventionally as p = 5, we have limited deviation from the given Baire 
array display data.

\section{Conclusions}

The Baire distance, simultaneously inducing an ultrametric as well as a metric,
is a longest common prefix metric.  It is just of interest to note that data 
transmission is commonly supported in computer networks through, as it is termed, 
longest prefix matching \cite{erdem}.   

In this work, we have approximated any particular set of decimal data by 
a representation in any number basis that is from 2 (i.e.\ binary, prime), 
3 (i.e.\ ternary, prime), 4, 5 (prime), 6, 7 (prime), 8, and 9.  Our methodology 
includes iterative optimization that is used for quantization.  Such iterative
optimization is not guaranteed to provide the global 
optimum, yet nonetheless with a fixed number of iterations, a practical sub-optimal
outcome is obtained. 

The relevance of bases other than 10, or 2 for binary arithmetic, can be addressed
in terms of application domain.  Following discussion of number base (and also 
scale) invariance, Hill \cite{hill} briefly comments on octal (base 8) and 
hexadecimal (base 16), as well as binary, systems for computation.   It can 
also be recalled that a ternary, 
i.e.\ base 3, computer was developed and built by Nikolay Brusentsov in 1958, and 
remains a technology, i.e.\ having a ternary rather than a binary basis for 
computers, that is still the subject of interest \cite{jones}.  

The ability to map, through approximation, between number systems, that we have 
developed here, is valuable, firstly, to minimize storage, and, secondly, to manage 
the data representation.  Such data representation in data analytics is 
for decision support, or, alternatively expressed, supporting actionable data.  

Our work supports the following insightful statement.  
In his book \cite{simon} (p.\ 2015), Herbert Simon, 
the Nobel prize-winner (1978, Economics), noted the 
following, in the chapter entitled ``The architecture of complexity: hierarchic
systems'':
``How complex or simple a structure is depends critically upon the way in which
we describe it. Most of the complex structures found in the world are enormously
redundant, and we can use this redundancy to simplify their description. But to
use it, to achieve the simplification, we must find the right representation.''

\end{document}